\documentclass[12pt]{article}
\usepackage{psfig}

\newcommand{\lqcd}{\ifmmode \Lambda_{QCD} \else $\Lambda_{QCD}$\fi}
\newcommand{\ra}{\rightarrow}
\newcommand{\etal}{{\it et al.}}
\renewcommand{\bar}[1]{\overline{#1}}
\newcommand{\VEV}[1]{\left\langle{#1}\right\rangle}

\newcommand{\ket}[1]{\,\left|\,{#1}\right\rangle}
\begin{document}

\bigskip
\begin{flushright}
SLAC--PUB--8285 \\
December 1999
\end{flushright}
\bigskip

\begin{center}
{\large Hadronic Effects in Two-Body $B$ decays}
\footnote{Support by Department of Energy 
Contract DE-AC03-76SF00515.}\\[3ex]
Helen Quinn \\[2ex]
{\it Stanford Linear Accelerator Center} \\
{\it Stanford University, Stanford, CA 94309}
\end{center}

\begin{center}
Abstract
\end{center}
\baselineskip 16pt
In these lectures I discuss the impact of soft hadronic physics on predictions 
for $B$ decays.  Unfortunately our tools for calculating these effects are limited; 
even after the use of the best available tools the resulting theoretical uncertainties 
are difficult to delimit, and can obscure tests for the presence of 
beyond-Standard-Model physics. The first lecture reviews what tools are 
available, the second reviews in more detail two examples of how these tools 
can be used.

\vfill
\begin{center}
Lectures presented at the XXVII SLAC Summer Institute of Particle Physics \\
``CP Violation---In and Beyond the Standard Model'' \\
SLAC, July 7-16, 1999
\end{center}
\vfill

\section{Lecture 1---Tools}

\subsection{What is the problem?}

In these lectures I will follow the notation and definitions given by
Yossi Nir in his lectures.\cite{yossitalk} (For another excellent set of
review lectures on CP Violation, including detailed references to the
original literature see lectures by A. J. Buras \cite{burasrev}.  
For a recent book also covering this topic in detail see ``CP Violation" 
by G. Branco \etal\ \cite{CPBook})

The physics of $B$ meson decays is governed by weak decay processes.  Weak
decays and any hard QCD effects are calculable by perturbation theory
methods, but soft QCD effects not are directly calculable.  Such effects
are inevitably part of the meson decay process; they define the internal
structure of mesons, the branching fractions to few and many-body
channels, and the interactions between final-state hadrons once they
have formed.  Their impact can mask our ability to relate measurements
to underlying Standard Model (CKM) parameters.  This problem is a
familiar one, it is not new in $B$ physics; in fact it is a much worse
problem for lighter meson decays.  The larger $B$ mass makes some of the
physics more calculable, but even in the limit of extremely large $B$ mass
there would be some work to do to deal with soft QCD effects.  

Hard and soft QCD effects are separated by the scale of the momenta
compared to the parameter \lqcd.  This is the scale at which the strong
coupling constant $\alpha_s$, as defined perturbatively, becomes
infinite.  Physically this scale sets the size of hadrons.\footnote{The
scale \lqcd is usually defined as the scale that determines the $q^2$
dependence of the QCD coupling at high energy; in leading order
$\alpha_s(q^2) = 12/[(33-2N_f)ln(q^2/\lqcd^2)]$ where $N_f$ is the
number of quark triplets.  This scale then defines where the
perturbative coupling becomes infinite, which is clearly well below the
scale at which perturbation theory is no longer reliable.  The physical
phenomenon associated with the growth of the coupling at short distance
is confinement, and one physical manifestation of that phenomenon is the
size of hadrons.  It is in this sense that \lqcd defines the size scale
of hadrons; the two scales are not numerically equal but are related
quantities.  The relationship cannot be calculated perturbatively, but
can be explored in lattice calculations.} Any freely propagating quark
or gluon with momentum small compared to this scale is a fiction---such
particles are not observed because of confinement.  Said another way:
in this regime QCD perturbation theory is not meaningful and nor are
Feynman-type diagrams, which are after all just a short-hand for
perturbative calculations.  Any time you see a line in a diagram for a
low-momentum quark or gluon you should be suspicious.  In reality any
such line comes dressed with a multitude of soft gluon emission and
absorption processes, and also additional soft quarks and antiquarks.
This part of QCD physics is not perturbatively calculable.  To
incorporate its very real effects we must resort to other tools.
Conversely, for quarks or gluons with momenta large compared to the
scale \lqcd \ QCD perturbation theory is an effective and accurate tool.

``Hadronic effects'' in my lecture title refers to the soft part of the
physics.  In my first lecture I will review what tools are available to
treat the problem and briefly comment on the uses of these tools. For some further
 discussion of some of the topics that I treat rather briefly here see  
A. F. Falk \cite{chap2}. In 
the second lecture I will turn to a few specific examples that
illustrate in more detail how these tools can be used.  Even with the
best available tools some residual uncertainties about the impact of
soft physics remains.  The term ``theoretical uncertainty'' is used here
to characterize impact of this poorly-calculated physics on the
extraction of well-defined parameters such as the elements of the CKM
matrix.  One unfortunate consequence of these uncertainties is that they
can mask possible new physics effects, as they obscure the relationship
between the data and clean Standard Model predictions.

The goal is then to minimize the parts of the calculation affected (or
should I say infected?) by these uncertainties.  In addition one hopes
that, eventually, comparison of data and calculations for many channels
can provide some confidence in the reliability with which the residual
uncertainties can be estimated.  However it is important to remember
that the estimates of these uncertainties are just that, estimates.
They may be based on nothing more than a particular theorist's gut
feeling about the subject.  The models and approximations used are often
simply not well-controlled enough for one to know how big the
corrections might be.  It is not justifiable to treat these estimated
uncertainties as if they were statistical errors.  This is often done;
procedures such combining these uncertainties in quadrature and quoting
probabilities for a deviation of twice the theoretical error as if these
were statistical standard deviations are all too common.

\begin{figure}[htb]
\begin{center}
\leavevmode
\psfig{figure=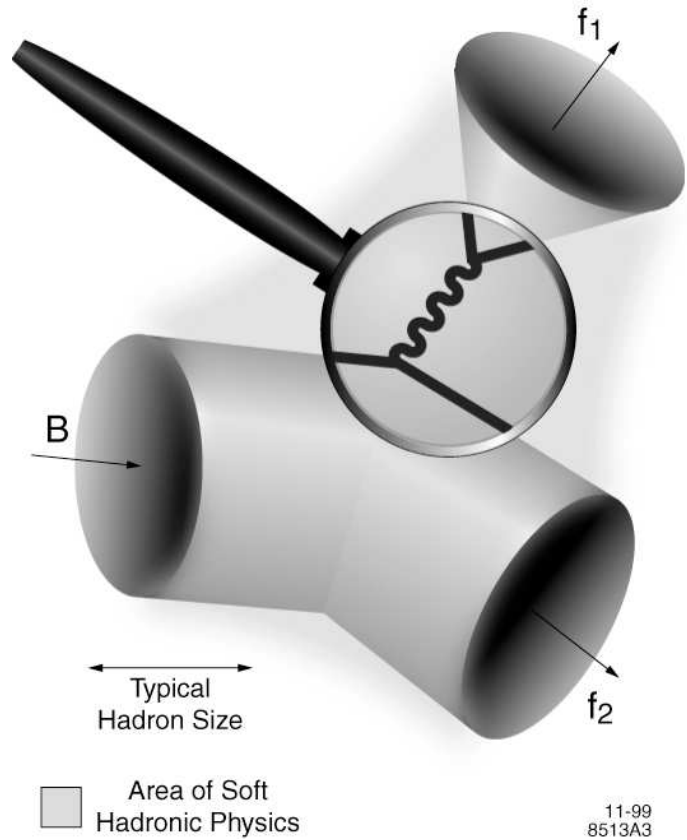}
\end{center}
\caption[*]{The hard and soft regions of a typical $B$ Decay.  The shaded area 
contains many (undrawn) soft gluons and quark-antiquark pairs, only the 
hard process (within the magnifying glass) is perturbatively calculable.}
\label{fig1}
\end{figure}

Figure \ref{fig1} indicates why the soft physics is usually unavoidable.  We can
use perturbation theory to calculate the part of the diagram within the
inner magnifying lens, that is the short distance parts.  The weak decay
is short distance because the mass of the decaying $b$-quark is small
compared to the $W$-mass, so the $W$ is highly virtual.  At the same time
(and here the difference with lighter mesons appears), the $b$-mass it is
heavy enough that the produced quarks in general have momenta which are
large compared to \lqcd.  In addition hard gluons exchanged between
these particles can be included perturbatively.  However the initial and
final hadron wave-functions, the quantities that describe these hadrons
in terms of their quark content, are not perturbatively known, nor do
they contain only hard quarks.  Even if we take the simplest possible
picture of a $B$-meson as a static heavy $b$-quark surrounded by some
wave-function distribution for the light quark, that light quark has a
typical momentum set by the size of the meson and hence, by definition,
of order \lqcd.  In most calculations of few-body decays this
``spectator'' quark (so-called because it does not participate in the
weak decay except in the case of annihilation diagrams where it is
clearly no longer a spectator), is assumed to hadronize as a valence
quarks of one of the final mesons.  This certainly does the book-keeping
of charge etc. correctly, but it gives a deceptively simple diagrammatic
picture.  Such a quark cannot be included in the hard or short-distance
part of the calculation; any estimates that depend on treating it as a
freely propagating particle are at some level suspect. 

In making the division between hard and soft physics an arbitrary and
unphysical scale $\mu$ is introduced into the problem.\cite{wilsonope}
This scale must be chosen to be large compared to \lqcd \ but is
otherwise unconstrained.  As is usual in QCD calculations one ends up
with terms of the form $\alpha_s(\mu)ln(k m_b/\mu)$ where k is some
number (probably of order unity) and the scale $m_b$ enters because it
is the quantity that defines the scale of momenta flowing in the hard
quark lines.  In order to avoid having this logarithm be large, it is
convenient to choose $\mu$ of order $m_b$, provided that does not make
$\alpha_s(\mu)$ too large.  Here is where the $B$ system theoretical
analysis is in much better shape than that for charm decays or
especially for $K$-decays.  The fact that $m_b$ is large compared to
\lqcd \ makes the hard/soft division a relatively clean business in $B$
physics.

In these two lectures I will confine my attention chiefly to two body
(or quasi-two-body) decays, for the sake of specificity.  The problem of
dealing with soft hadronic physics effects is not unique to calculations
of two body decays, nor are the general statements made below about
methods and symmetry limits special to those decays.  Many of the
general approaches I mention here also have some applications for
inclusive processes and for many body decays.  My intent here is not to
teach you to use any of the tools that I discuss, but rather to make you
aware of them and of their uses, and their limitations.  To actually
learn to use these tools and approximation methods requires more time
than we have available in these two lectures. 

As was demonstrated in Yossi Nir's lectures\cite{yossitalk}, two-body
and quasi-two-body decays to states which are $CP$ eigenstates are of
particular interest in neutral $B$ CP violation studies.  Such states
occur only for the case of two pseudoscalars, or one higher spin
(typically spin 1 is studied) and one pseudoscalar, because for both
these cases the decay of a spin-zero $B$ can give only one possible
relative angular momentum for the two produced particles.  Hence a state
of definite CP is produced.  For two higher spin particles, even when
the particle content of the state is CP-self-conjugate, the $B$ decay
produces an admixture of CP-even and CP-odd final states because both
even and odd relative angular momenta between the two produced particles
are allowed.  In many cases such systems can be separated into states of
definite CP via angular analysis of the decays of the two quasi-stable
``final state'' particles.\cite{angan} Then methods similar to those
discussed here for the simpler modes can be applied, once sufficient
data is available.  Without this separation a ``dilution'' or
cancelation effect occurs in the measured asymmetry; the CP-odd states
contribute the same asymmetry as the CP-even ones except for an overall
sign, so the two contributions partially cancel each other.

Methods for extracting CKM parameters from asymmetries in production of
inclusive final states with a particular CP-self-conjugate quark content
have been suggested.\cite{inclnocharm} These depend on estimates of the
CP-even and CP-odd fractions of the decay final states.  Such estimates
are made at the quark level.  They are reliable for the total inclusive
rate because hadronization, being a strong interaction process, respects
CP symmetry.  Typically they suffer from large hadronic uncertainties
once any cuts are introduced.  Such cuts are unavoidable; they are
needed either to define experimental apertures or to discriminate data
from backgrounds.  I will not discuss such methods further here.

\subsection{Scales, Exact Limits and Expansions around them.}

One of the things that makes the physics of $B$ decays complicated is that
many scales can play some role in the problem.  Roughly in order of
increasing size these are	 
$$m_u, m_d, m_s, \Lambda_{QCD},m_c,m_b,\mu$$
where the scale $\mu$ is an unphysical parameter introduced in QCD
marking the division between hard and soft QCD effects calculations
while \lqcd \ is the scale that defines the running of the coupling
in QCD.

For any significant hierarchy in these scales it is instructive to
pursue the limit in which either the small scale is taken to zero, or
the larger one to infinity.  For example, since $\lqcd/m_b \ll 1$,and
$\lqcd/m_c<1$ the heavy quark limit $m_c/m_b$ fixed, $m_b=\infty$ is a
useful approximation to the real world.\footnote{For some purposes the limit 
$m_b - m_c$ fixed, $m_b \ra \infty$ may be convenient to consider; 
it is important to recognize that there are subtle differences between 
these two variants of heavy quark limits, and to be aware which is used 
for a particular argument.}    It is useful for two reasons.
First the theory has additional symmetries which provide exact
constraints in this limit.  These constraints can be used to limit or
relate various model parameters by requiring that the model have the
correct limiting behavior.  Second, one can calculate corrections to
this limit as a power series in small quantities, namely ratios
$\lqcd/m_b$ and $\lqcd/m_c$.  This is called the heavy quark expansion.
One has good control over the sizes of neglected corrections and hence
over theoretical uncertainties due to these corrections.  Unfortunately
the second ratio, $\lqcd/m_c\approx 1/3$ is not so small in the real
world; quantities where such terms are not suppressed have significant
corrections to the limiting behavior.  Cases where the leading
correction is second order in this ratio are particularly attractive for
this approach.
Working down the scale hierarchy the following approximations and limits
can be considered:

\begin{tabular}[t]{rll@{\extracolsep{.1in}}}
$(m_s-m_d)/\lqcd<1$  &    Limit: $m_u=m_d=m_s$ & SU(3) Invariance \\
$(m_d-m_u)/\lqcd \ll 1$ &   Limit: $m_u=m_d$  & Isospin Invariance \\
$m_u/\lqcd \ll 1,m_d/\lqcd \ll 1$ &   Limit: $m_u=0, m_d=0$ & Chiral Invariance.
\end{tabular}
\bigskip

\noindent
Each of these limits can be useful in restricting uncertainties in
hadronic physics effects by introducing constrained parameterizations
with somewhat controlled corrections.  Examples will be discussed in
more detail below, and in the second lecture.  One point of caution:
sometimes the interplay of more than one of these scales can limit the
effectiveness of such expansions.  Terms which might be treated as small
because they contain inverse powers of a large mass cannot be
disregarded if at the same time they contain inverse powers of a small
mass.

Many of the methods for estimating matrix elements or form factors do
not introduce any explicit $\mu$ dependence in them. Thus, at best, the
estimates can be valid at only one value of $\mu$.  Often we have no
good arguments to choose that scale.  It is not uncommon for theorists
to characterize the uncertainty introduced by this error in matrix
element calculations by looking at the variation of the result over the
range from $m_b/2 \leq \mu \leq 2 m_b$.  This choice of range has no
theoretical justification.  In some model calculations the natural scale
for the model is a light hadron mass scale, too small a scale to be
acceptable from the QCD point of view.  Because of this mismatch between
the scale at which the model estimate of the matrix elements can be made
and the plateau region of the coefficient calculation it is difficult to
characterize the size of the uncertainty in calculations that depend on
such models.  Methods such as lattice calculation where the matrix
element calculation does have explicit scale dependence give much better
hope for eventual results with well-controlled uncertainties.\cite{latticereview}

\subsection{Heavy Quark Limit}

This limit is most useful in the context of decays $B \ra D X$,
particularly the semi-leptonic processes; for example it provides
important control over the theoretical uncertainties in the extraction
of $V_{cb}$.  The best cases are those where the leading correction is
quadratic in the quantity $\lqcd/m_c$, since this ratio is not small
enough for terms proportional to a single power of it to be a small
correction.  For channels with no final state charm particles one can
use the heavy quark limit to relate $B$ decays to corresponding $D$
decays, for example extracting the behavior of form factors for $B$
decay from those measured in the $D$ decay case.  The accuracy of this
approach is limited, both by the accuracy with which the $D$ decays are
measured and by $\lqcd/m_c$ corrections.  There is a large literature on
the subject of heavy quark limit calculations, I will not discuss these
methods further here.\cite{heavyquarkreview}

The heavy quark limit is generally applied for hadronic $B$ decays only in
combination with the factorization approximation.  In the cases $D X$
 it has been shown that factorization is
valid in the heavy quark limit for a particular kinematic region.\cite{dugan}  
For charmless decays the combination of
the two methods adds uncontrolled theoretical uncertainties.

\subsection{Isospin}

Isospin analysis is a useful tool in some $B$ decays, principally for
its role in separating gluon-mediated penguin contributions from
tree-diagram contributions (see Yossi Nir's lectures for definition of
these two types of diagrams).  The crucial point is that gluons have
isospin zero which limits the isospin amplitudes to which they can
contribute.  The details of how the isospin information is used depends
on the channel.  I will review this in some further detail for a couple
of channels below and in my second lecture.

For this young an audience it is probably necessary to start a
discussion of isospin analysis by defining what is meant by isospin.
Isospin is an SU(2) algebra in which the up and down quarks are treated
as two identical members of a doublet.  Note this strong interaction
doublet is similar to, but not the same as, the weak SU(2) (sometimes
also called weak isospin) doublet which pairs the up quark with a linear
combination of down-type quarks
\begin{equation}
d \cos(\theta_{12})\cos(\theta_{13})+s \sin(\theta_{12})\cos(\theta_{13})+ b \sin(\theta_{12})\sin(\theta_{13}).
\nonumber
\end{equation}

Isospin is a symmetry of the strong interactions but not of electroweak,
which clearly distinguish quark charges and flavors.  It is also broken
by quark mass terms.  Historically the name isospin came about because
physicists were familiar with the SU(2) algebra as the algebra of spin,
and from the relationship of the multiplets of this symmetry to the
isobars of nuclear physics (nuclei of equal A).  From a modern
perspective we can understand that hadrons form approximately degenerate
isospin multiplets with mass differences small compared to the average
mass of the multiplet because most hadron masses are dominated by
$\lqcd$.  The up-down quark mass difference is small on this scale, even
though their mass ratio is far from 1. The exception is that
pseudoscalar octet masses scale as $\sqrt{m_q \lqcd}$ \footnote{This scaling
follows from the pseudo-goldstone nature of the pseudoscalar mesons and
the PCAC (partially conserved axial current) relationships such as
$m_{\pi}^2f_{\pi}^2 = ({m_u+m_d})\VEV{\bar \psi \psi}$, since $f_\pi\propto
\lqcd$ and $\VEV{\bar \psi \psi}\propto \lqcd^3$ are both quantities whose
scale is defined by QCD confinement physics.} and thus the effect of
quark mass differences can give larger isospin breaking in this multiplet.

Isospin-breaking can also be significant in the neutral meson states.
Ideally the two neutral quark-antiquark states have $I=0$ and
($I=1,I_3=0$):  $ \eta_{ud}$ and $\pi_0$ for the pseudoscalars, $\omega$
and $\rho $ for the vectors.  In actuality, because the up and down
quark masses are not identical, the mass eigenstates have small
admixtures of the wrong isospin state.  This can lead to important
contributions that are neglected if the physical particles are treated
as having a definite isospin.  \cite{ispinbreaking}  (The notation
$\eta_{ud}$ also serves to warn that, for the pseudo-scalars, the
strange-antistrange combination is also mixed into the physical $\eta$
particle; $\eta_{ud}$ means that combination of $\eta$ and $\eta^\prime$
with no strange quark part.)

The photon and the $Z$ each couple to up and down quarks with a
well-defined ratio of I=0 and I=1 couplings, for both vector and, in the
case of the $Z$, axial vector couplings.  These couplings are usually
written in terms of coefficients $g_1^X,g_2^X$ for coupling to $u$ and
$d$ quarks respectively, with superscripts $X=V, A$ for the vector and
axial vector couplings respectively.  The combination $(g_1^X\pm
g_2^X)/\sqrt{2}$ are the definite isospin couplings.  This relationship
between coefficients gives a relationship between the amplitudes of
definite isospin for a given $Z$-mediated or photon-mediated process if
final state interactions are neglected.  The final state interactions
introduce corrections, including complex phases from absorptive parts,
which are in general different in the different isospin states and
cannot be calculated from first principles---that is without further
assumptions.

Since photons and $Z$ bosons have $I=1$ as well as $I=0$ couplings to
quark-antiquark states electroweak penguin effects cannot be removed by
the same isospin analysis that eliminates QCD penguin effects.
\cite{qedpenguins} Their impact varies from channel to channel, but must
be considered.  This limits the usefulness of isospin in removing
hadronic uncertainties in the extraction of CKM parameters from CP
violation in some weak decays.  However in many channels the electroweak
penguin effects can be shown to be small.  Then the uncertainties that
they induce in the extraction of CKM parameters are likewise small.

As an example of how isospin enters in $B$-decays let us consider the
decays based on the quark process $b \ra u\bar u d$.
\cite{gronaulondon} These three final quarks can have either $I=1/2$ or
$I=3/2$, thus we can label the quark transition as $\Delta I =1/2$ or
$\Delta I =3/2$.  The additional (spectator) quark (and hence the
charged $B$ and $B_d$) are an isodoublet.  Thus, combining this initial
isospin with the transition isospin $\Delta I$, we find four
possibilities $\Delta I =1/2, I_f=0$; $\Delta I =1/2, I_f=1$; $\Delta I
=3/2, I_f=1$ and $\Delta I =3/2, I_f=2$ for these decays.  The gluonic
penguin can contribute only to the first two cases, because the gluon
couples only to the $I=0$ combination of quarks $u\bar u + d \bar d$.
Hence gluonic penguin contributions have $ \Delta I =1/2$ only.  Any
pure $I=2$ contribution is thus unaffected by gluonic penguin contributions.  
Up to corrections from electroweak penguins, it has the property 
$\bar A_2/A_2= 1$ in the Standard Model.  Thus, if this contribution 
can be isolated, it can provide a relatively clean estimate of the related 
CKM parameter in channels where the
electroweak penguin effects can be demonstrated to be small relative to
the dominant terms. 

 Another reason to arrange the calculation in terms
of isospin amplitudes is that final state interactions mix states of
different charge structure but, since they are strong interaction
effects, do not change isospin.  Let us expand in the basis of strong
interaction eigenstates $\ket{i^I}$, for which the scattering matrix is
diagonal.  The diagonal strong interaction scattering matrix contains an
independent strong phase for each entry
\begin{equation}
\VEV{j^I|{\cal H}|i^I} = \delta_{ij}e^{2i\delta^I_i}.
\end{equation}
The eigenstates have definite isospin, but include both two-particle and
many-particle components.  Thus more than one eigenstate $\ket{i^I}$ exists
for each isospin $I$.

The kinematic structure of each operator is different, thus the states
of given isospin produced from the $B$ by two different operators are,
in general, different linear combinations of the strong interaction
eigenstates; we write $\VEV{i^I|{\cal O}_j|B} = x^I_i({\cal O}_j)$.  The
rescattering effect introduces the square root of the scattering
matrix\cite{fsi}; heuristically one sees this by noting that the process
is not going from an in state to an out state, but starting ``in the
middle'' from a pointlike local superposition and evolving to an out
state.  Finally, to consider a given two-body final state $f^I$ one
needs the overlap $\VEV{f^I|i^I}= a_i^I(f)$.  Thus one can write
\begin{eqnarray}
 A^I({\cal O}_j,f)&=& \Sigma_i <f^I|i^I>e^{i\delta^I_i}\VEV{i^I|
{\cal O}_j|B}  \nonumber\\
&=& \Sigma_i a^I_i(f)e^{i\delta^I_i}x^I_i({\cal O}_j).
\end{eqnarray}

This expression is not very useful since, in general, we cannot
calculate any of the quantities in the right-hand side.  However, it
does serve to destroy a couple of myths that appear now and then in the
literature.  The first is that the only effect of rescattering is to
introduce a phase in the isospin amplitudes.  The second is that the
strong phase for an amplitude with a given isospin is the same
independent of the operator.  A little playing with the above
expression, say for the cases where there are just three strong
eigenstates, will show that neither of these statements is true in
general.  One sees that they would each be true if there were only a
single strong eigenstate excited for each isospin, or if the two-body
state of definite isospin were by itself a strong interaction
eigenstate.  (In general, neither of these conditions is true.)
\footnote{The misperception that just one state and hence one phase
exists for each isospin is perhaps a holdover from low energy isospin
physics, where it is true because the multibody channels are
kinematically excluded.} 

\subsection{SU(3) Symmetry}

This is another approximate strong interaction symmetry, very much like
isospin except that in addition to equal mass up and down quarks the
symmetry limit requires the strange quark mass to be degenerate with
them.  Since the ratio $(m_s-m_d)/\lqcd$ is not so small, SU(3) breaking
effects can be large.  In $B$ decays the most common use of SU(3),
beyond its isospin subgroup, is the application of results due to
another SU(2) subgroup of the SU(3), traditionally called U-spin.
U-spin treats the $s$ and $d$ quarks as a doublet of identical particles.
For example, it relates rates where pions are replaced by kaons, and/or
$B_d$ by $B_s$.  \cite{suthree}

In the factorization approximation, for any quantity where an axial
current produces a pseudoscalar meson the SU(3)-breaking effect is
known, it is the ratio $f_K/f_\pi$, which is measured to be $1.22 \pm 0.01$.
For the vector current producing a vector meson the relevant correction
factor is $F_K/F_\pi$.  Similar corrections occur for transition matrix
elements.  These corrections provide, presumably,  a
good first estimate of SU(3) breaking corrections, though there may be
further corrections due to differences in final state scattering effects
for the two different mesons.  At the $B$ mass scale it is reasonable to
assume that these are small corrections.  However for many other
contributions the use of the ratio $f_K/f_\pi$ (or $F_K/F_\pi$) to
estimate the SU(3) breaking is not justified even in factorization
approximation, and large theoretical uncertainties remain.  In some
calculations the two SU(3)-related amplitudes for these cases are
allowed independently parameterized magnitudes and the SU(3) symmetry
approximation is applied only to identify their strong phases.
\cite{suthreephases} Once again the corrections to this
approximation are expected to be small at the $B$ mass.  However I do not
know how to quantify the expected size of ``small'' effects due to SU(3)
breaking of the strong-rescattering phase relationships.  In any
particular case one can test the impact of relaxing this constraint by
looking at how the fit for the CKM parameters of interest change with
the difference between the two strong phases, but no clear statement
prescription for what would be a ``reasonable range'' of phase
differences to allow in such a treatment can be given.

\subsection{Chiral Symmetry}

The chiral limit and chiral perturbation theory are based on the
approximation that the up and down quarks are massless in which case the
pion is a Goldstone boson.  This leads to an expansion of amplitudes for
the production of an additional soft pion in terms of the amplitude
without that pion and correction terms which occur as powers of the
momentum of the soft pion scaled by $\lqcd$.  (This scaling defines what
is meant by soft in this context.)  While this method has some uses in
$B$ physics calculations \cite{chiral}  it is not a
useful tool for the treatment of two body hadronic decays, since the
pions produced in such decays are not soft.  I will not discuss chiral
expansions further in these lectures.

\subsection{QCD Sum Rules}

These are conditions derived from the analytic structure of QCD
perturbation theory.\cite{qcdsumrules} Sum rules typically relate
certain matrix elements or derive constraints on their kinematic form in
particular limits.  Such constraints are useful in limiting the
arbitrariness of models, for example those for form factors in
semi-leptonic decays.  The BaBar Physics Book contains an appendix which
discusses this subject. I will not treat it further in these lectures.

\subsection{Lattice calculation of matrix elements}

Ideally we need a method for calculating the long distance
contributions, that is the matrix elements, that correctly includes all
soft physics.  This would also give the correct sensitivity to the
hard-soft division scale $\mu$.  The method with the best hope of doing
this is lattice calculation.\cite{latticereview} QCD sum rules can also
be used to extract information about certain properties of form factors,
but are not powerful enough to calculate the matrix elements themselves.
Unfortunately, for most the cases of interest here, the same thing must
be said about the lattice calculation of matrix elements, at least at the current state 
of the art.

For two-body $B$ decays these matrix elements are three-point functions
connecting the initial $B$ to the two final-state particles.  In
actuality what is calculated on the lattice so far is a less-demanding
two-point function, where one of the final particles has been ``reduced
in''.\cite{LSZ} It thus appears in the operator that is evaluated, rather
than as a final state particle.  This removes all sensitivity of the
calculation to final state interaction phases, which are one of the
major issues for CP-violation physics.\cite{sharpe} Furthermore, most of
the relevant lattice calculations have so far only been made in the
``quenched approximation'' ---which means in the approximation of
suppressing any virtual quark-antiquark-loop contributions.  As with
experiments, lattice calculations then have a statistical uncertainty of
their result and in addition non-statistical (or systematic)
uncertainties arising from these various simplifying approximations.
The former are readily estimated and clearly given in lattice results,
the latter are hard to estimate and hence again significant theoretical
uncertainties remain in most cases.

Where an unquenched calculation exists results are sometimes
significantly different from unquenched results for the same quantity.
We have no good understanding of how to quantify these differences prior
to making the more difficult unquenched calculations.  A growing number
of unquenched calculations are appearing, but as yet no true three-body
calculations.  Again there is a large literature on this subject and I
do not the time (nor the expertise) to cover it in
detail.\cite{latticereview}

There are a number of quantities relevant to the extraction of CKM
parameters from $B$ physics for which the lattice calculations are in
much better shape than for the three body matrix elements discussed
above.  For quantities such as $F_B$, and many the various $B_i$
parameters (parameterizing the ratio of true matrix element to vacuum
insertion approximation results for the QCD operators ${\cal O}_i$)
unquenched calculations are beginning to be feasible.  Reliable values
(with uncertainties in the few percent range) are expected for most of
these quantities within the next few years.

\subsection{When are these methods useful?}

I have summarized a fairly large ``bag of tricks'' for dealing with
hadronic effects.  Remembering Feynman's dictum that if you have one
good method you don't need any others, the length of the list alone
should give you an idea of the state of the problem!  The applicability
and efficacy of each of these methods varies from channel to channel.
In the best cases we do not need any of them, because, as Yossi
explained, when amplitudes with only a single weak phase dominate a
decay, as is the case for the channel $J/\psi K_S$, the hadronic
amplitudes cancel out in the ratio that defines the CP asymmetry.  Then
none of the uncertainties in calculating the matrix elements matter.
Such a mode gives the cleanest relationship between a CKM matrix element
phase and a measured asymmetry.  Conversely the problems are worst when
the same channel receives two comparable-magnitude contributions, say
from suppressed tree diagrams and from penguin diagrams, or from two
different penguin diagrams in a channel with no tree contributions, and
the two contributions enter with different weak-phases, that is with
different CKM matrix element coefficients.  In each such case the
relative strength and the relative strong phases of the two
contributions affect the relationship between the measured asymmetry and
any CKM parameter.  One must then use whatever tools are available to
try to make estimates of these effects, and equally important, to
constrain the uncertainties in these estimates.

\subsection{Approximations that do not come from exact limits}

In many cases the methods described above are not sufficient to obtain
all the desired information.  When this is the case one is forced to
resort to less-controlled approximations, which generally have some
intuitive model as their underpinning.  Such methods are very useful,
for example to obtain estimates of the expected branching fraction for
various channels.  The most commonly used approximation is that of
factorization, which I will discuss shortly.  It is difficult to obtain
any good estimate of the theoretical uncertainties introduced by such an
approximations.  Thus it is very difficult to find convincing evidence
for non-Standard-Model contributions from any conflict between such
estimates and measured results.  However they are part of the standard
toolkit for calculating $B$-decay processes and so are worth mention here.

\subsection{Factorization}

This approximation starts from the operator product expansion and
provides an estimate of the matrix element of the local four-quark
operators.  One takes any such operator and finds any possible
Fierz-rearrangement that groups the four quark fields into two that can
create one of the final-state hadrons from a vacuum state, and two 
that describe a transition
matrix element from the $B$ to the other final state hadron.  All final
state interactions between the two hadrons are ignored, as are any
operators that cannot be arranged in this way.  This is a very useful
approximation as it allows a few-parameter model to describe many
two-body decays, using transition matrix elements
measured elsewhere, for example in semileptonic decays.

The idea behind this ansatz is that the region of the phase space
where the two-body final state is most likely to be produced is that
where two quarks that form a meson are produced moving roughly together
and in a color-singlet combination.  Since the operator that produces
them is local, the state so made is a local color singlet state.  Hence,
unlike a real finite-sized hadron, it has a very small strong
interaction cross section with the other quark-antiquark system.  Since
the two systems are rapidly moving apart, they are far separated from
it before the local state has evolved into its final finite-sized
configuration as a hadron.  Thus it can be expected that no significant
strong interaction rescattering occurs between the two mesons so formed.
This ``color-transparency'' argument is attributed to
Bjorken.\cite{bjfactorization}

When the two quarks that have the right flavor and tensor structure to
form the single meson are not automatically in a color singlet state the
color transparency argument is less immediately obvious.  Effectively
the requirement that the meson is formed projects out the color singlet
part of the $\bar q_\alpha \Gamma_i q'^\beta$ operator (here $\Gamma_i$
denotes some gamma-matrix structure and $\alpha$ $ \beta$ are color
indices).  The color counting then gives a suppression of $1/N_c$ since
the ``color-allowed'' contribution
\begin{equation}
 \Sigma_{\alpha}\VEV{m_1|\bar q_\alpha \Gamma_i q'|0}
\Sigma_{\beta} \VEV{m_2|\bar q_\beta \Gamma_i q'^\beta|B} \propto  N_C^2
\end{equation}
 whereas the contribution
\begin{equation}
\Sigma_{\alpha}\Sigma_{\beta}\VEV{m_1|\bar q_\alpha
\Gamma_i q'^\beta|0} \  \VEV{m_2|\bar q_\beta \Gamma_i q'^\alpha|B}
\propto  N_C \ .
\end{equation}
 This is  the ``color-suppressed'' factorized contribution.

If the argument for neglecting final state interactions is rephrased in
the language of strong interaction eigenstates given in the isospin
section above, it looks much less attractive.  As best I can see, it
seems to say that the operators excite a linear combination of strong
interaction eigenstates each of which gets a strong phase from
rescattering, but in such a way that their vector sum is unchanged.
(Another option, that looks even less plausible, is that the $B$-decay
forms only a single strong interaction eigenstate involving any two pion
component, and that that state has zero rescattering phase.)The general
formalism instead suggests that configurations where the two quarks that
make the final meson are not produced traveling together can
contribute, via rescattering, to the two-body final state, even when
naive expectations say that is unlikely.  This contribution may indeed
be small, but we cannot say how small.  Our intuition rejects this
possibility just because we know that for any given many-body state the
probability of rescattering to two pions is typically small.  However,
at the $B$-mass, the cross section for two pions in an $s$-wave to scatter
into many pions is not expected to be small.  Thus the inverse process
must also be possible for some configurations of the many particles.
The problem is that any way of making the exclusive two body final state
is suppressed, either because it involves a small corner of the
four-quark phase space where two quarks happen to move together or
because it involves a many particle to two particle rescattering.
Intuition is generally a remarkably poor guide to discovering which of
two unlikely events is more likely.  I make this comment just to show
how little we actually know---and that models can seem quite plausible
in words but have little calculational basis.  It is not that I know the
color transparency argument is wrong---just that I know no way of
proving that it is right either.

There have been a number of papers devoted to the impact of final state
interactions, which are neglected in the factorization approximation.
Some approach the problem generally, others consider specific channels.
Some sample papers on this topic are given in the references. \cite{fsi2}

Recent work by Beneke, Buchalla, Neubert and Sachrajda
\cite{benekeetal} has introduced a more detailed study of how this
factorization idea plays out in a one loop calculation, and at leading
order in $\lqcd/m_b$ Their approach depends on certain assumptions, such
as the dominance of the simple quark-antiquark state in the composition
of the meson wave-function, compared to any contribution where
additional soft quarks and antiquarks play a key role.  It is not based
on a rigorous operator product starting point, even in the infinite
$m_b$ limit.  They find that there are certain additional contributions
that are ignored in the simplest factorization calculations, which means
there are more input parameters to be determined in their calculations
than in the usual  factorization approximation calculations.  However
once these contributions are added they find that final state
interactions are suppressed at the one loop level, because of
cancelations of the type one would expect from color-transparency 
arguments such as that given above.  They are currently in the process of
extending their study to the level of two-loops.

One problem with the factorization approach is that is gives no scale
dependence for the matrix elements.  Since the coefficients are scale
and renormalization-scheme dependent, naive factorization cannot be
precisely true except possibly at some particular scale, and in
conjunction with a particular choice of renormalization scheme.  A
common approach to this problem is to use the induced scale and scheme
dependence as an estimate of the theoretical uncertainty of the method.
However this is surely not a rigorous argument, firstly because the
answer depends on the range of scales allowed, and secondly because it
gives no estimate whatsoever of the contributions that are ignored in
the factorization approximation.  The best one can say is that this
dependence sets a lower bound on the theoretical uncertainty.  But of
course what we really need is an upper rather than a lower bound on
uncertainties.

\subsection{Quark Hadron Duality}

This set of theoretical buzz words has two basic versions---global
duality and local duality.  Global duality is the statement that when
averaged appropriately over some range of center of mass energies the
rate for a given process predicted by a quark level calculation must be
the correct result for the rate at the hadron level.  For certain
quantities such as the ratio of the hadronic cross section to the
$\mu^+\mu^-$ cross section in $e^+e^-$ scattering this can be
demonstrated to follow from the analyticity structure of the propagator
function $\Pi(s)$.\cite{pqw}

Local duality is the same idea applied at a given center of mass energy.
In $B$ decays we cannot vary the energy, it is the $B$ mass, so to
relate the quark quantities we know how to calculate to the hadronic
quantities we know how to measure we are forced to make this stronger
assumption.  There is no good justification for the truth of this
assumption, nor is there any good way to estimate the size of the
uncertainty it introduces.  Even within the assumption of local duality
there is a weaker and a stronger form.  The weaker assumption is to
apply duality arguments to calculate rates for a particular class of
inclusive decays, the stronger assumption is to rely on details of the
quark-level kinematics to predict the hadron-level properties.  In fact
at the end points and in resonance regions of the spectrum this last
approximation must be wrong, because quark kinematics does not know
about resonance widths and hadron masses, etc.  As soon as one goes from
a truly inclusive prediction to one that takes into account any
experimental acceptance cuts the predictions tend to become dependent
on this strongest form of the quark-hadron duality assumption, and the
theoretical uncertainties increase accordingly.

\subsection{Parameterized Amplitudes and Models}

Another way that one can proceed is to introduce parameters for each
diagram or each isospin amplitude.  One then obtains constraints by
relating the parameters describing similar contributions in different
processes, via symmetries such as isospin and SU(3).  Conversely one can
use models to calculate the value of the parameters for each type of
contribution.  Here the hope is that, with enough channels studied,
these parameterized amplitudes will eventually become sufficiently
constrained to be predictive.  The goal is that the estimates be
reliable enough to make relatively definite predictions about some of
the interesting quantities, and set relatively reliable constraints on
the theoretical corrections to a given calculation.  It is certainly
true that with enough data from enough channels we can begin to get a
better control.  Whether that control will become good enough that we
could unambiguously identify a non-Standard-Model contribution in
channels where more than one amplitude contributes remains to be seen.
The history of calculations of hadronic effects in $K$-decay processes, or
even $D$-decays, does not give grounds for optimism.  Here we are working
in a very different kinematic regime and the asymptotic freedom of QCD
and the heavy quark limit begin to work in our favor.  Time alone will
tell how well we can do.

\section{Lecture 2---Examples}

In this lecture I will review some examples where the tools of isospin
analysis and SU(3) discussed in the previous lecture may be useful.  I
will also make a few comments on the impact of hadronic effects in
extracting the magnitude of CKM matrix elements, such as $V_{ub}$.

\subsection{Isospin analysis for $b \ra u \bar u d$ channels}

\subsubsection{Two pions}

In the case of two identical particles in an orbital-angular-momentum
zero state (because they are two pseudoscalars coming from a $B$ decay)
the set of isospin amplitudes described for this quark content in my
first lecture ($\Delta I =1/2, I_f=0$; $\Delta I =1/2, I_f=1$; $\Delta I
=3/2, I_f=1$; and $\Delta I =3/2, I_f=2$) is reduced.  Bose statistics
requires a state of even isospin, so that the overall state is even
under the interchange of the two pions.  Hence the $I_f=1$ amplitudes
are all identically zero.  This means only two tree amplitudes, $\Delta
I =1/2, I_f=0$ and$\Delta I =3/2, I_f=2$, and only one penguin
amplitude, $\Delta I =1/2, I_f=0$, contribute.

Gronau and London\cite{gronaulondon} showed how a measurement of rates
for all three channels $B_d \ra\pi^+\pi^-$, $B_d \ra\pi^0\pi^0$, $B^+
\ra\pi^+\pi^0$ and their CP-conjugates, together with a time dependent
asymmetry measurement for the charged pions only, can be used to isolate
the weak phase of the $I_f=2$ contribution.  In principal, this method
provides a clean measurement of $\sin(2 \alpha$), where $\alpha$ is the
angle $\pi - \beta -\gamma$ in the unitarity triangle.  Unfortunately
the rates for all these channels are low,\cite{cleo} and the rate for
the difficult to measure $\pi^0\pi^0$ channel is expected to be even
lower.  It appears that the uncertainty of the measurement of this last
channel will render the method impotent to obtain a precise
result.\cite{babarbook} Put another way, for the foreseeable future the
experimental uncertainty on the neutral-pion measurement will be at
least as large as the theoretical uncertainty in the shift of the
measured charge-channel asymmetry from the simple form
sin($2\alpha$)sin($\Delta m t)$.

 \subsubsection{$\rho\pi$ channels and the Dalitz plot}

For $B_d\ra \rho \pi$ three channels contribute, namely the three
possible charge assignments for the $\rho$ and the pion, all decaying to
the same final state $\pi^+\pi^-\pi^0$.  However, by the arguments given
in the previous lecture, only two independent QCD-penguin amplitudes
exist.  One can take the three independent tree amplitudes to be one for
each charge channel and the QCD penguin amplitudes to be one each for
$I_f=0$ and $I_f=1$.  (If one plans also to use charged $B$-decay
amplitudes to three pions one additional tree amplitude enters; one must
measure both the three-charged and the two-neutral, one charged pion
final states before significant additional constraints are obtained in
this way.  The latter is more difficult experimentally, so I will here
discuss a study involving only the neutral $B_d$-decays to three pions.)

Five independent amplitudes, one CKM parameter and only three channels
looks a bit discouraging.  However Art Snyder pointed out to me an
important feature of the physics here that could be useful.  In some
regions of the Dalitz plot more than one of the three channels can
contribute.  Hence there might be information to be extracted from the
interference effects in the overlap regions.  Based on his suggestion we
made a preliminary study of this channel and found that this is indeed
the case.  The number of parameters to be fitted requires a large data
sample.  \cite{quinnsnyder} Further studies made as part of the BaBar
Physics workshop confirm this conclusion, and find that, as one might
expect, the inclusion of physics backgrounds from other resonances and
from non-resonant $B\ra 3\pi$ decays, as well as non-$B$ backgrounds
make things even more difficult.  However the analysis remains an
intriguing if distant possibility, so I will describe a little how it
works.

The amplitudes for the specific channel decays can be written
\begin {eqnarray}
A^{+-}& = &(T^{+-} +P_1+P_0) \nonumber\\
A^{-+}& = &T^{-+} - P_1+P_0 \\
A^{00} & = & T^{00}- P_0 \ .\nonumber
\end{eqnarray}
The assumption made in this approach is that each of the five
contributing tree and penguin amplitudes for $B_d \ra \rho \pi$ has an
independent but fixed ({\it i.e.}  not kinematically varying over the $\rho$
width) strong phase, along with a weak phase given by the Standard
Model.  The weak phase is different for the tree graph contributions and
for the dominant penguin contributions.  Further the weak phase of that
penguin contribution cancels the weak phase of the mixing.  Using the
Unitarity of the CKM matrix the sub-dominant penguin contributions can
be chosen to have the same weak phase as the tree amplitude; in all
further discussion of phase structure these contributions are assumed to
be included in the tree terms.\cite{pdgreview} (Note however that
in making numerical
estimates these two types of contributions must be considered
separately.)

The additional feature of this mode is that the full $B_d \ra
\pi^+\pi^-\pi^0$ is a sum of the three specific $\rho$-charge amplitudes.
It thus contains known kinematically-varying strong phases that arise
from the Breit-Wigner form of the $\rho$ resonances (more precisely stated
from the $\pi\pi$ scattering phases shifts in the $\rho$ resonance
region, which are parameterized by this form).  Thus the amplitude for
$B_d$ decay is given by
\begin{eqnarray}
A(B_d\ra \pi^+\pi^-\pi^0)  &=& f(k^+,k^0) (T^{+-} +P_1+P_0) \nonumber\\
 &+& f(k^-,k^0) (T^{-+} - P_1+P_0) + f(k^+,k^-0) (T^{00}- P_0)
\end{eqnarray}
where $f(k_i,K_j)$ is the Breit Wigner function
\begin {eqnarray}
f(k_i,k_j) &=& {cos(\theta)\over s-m_\rho^2 +i\Pi(s)}\nonumber \\
\Pi(s)&=&{m_\rho^2\over\sqrt{s}}\left({p\over
p_0}\right)3\Gamma_\rho(m_\rho^2)
\end{eqnarray}
where the $k_i$ are the momenta of the two pions, $s= (k_1+k_2)^2$, and
$\theta$ is the angle in the $\rho$ rest frame between $k_1$ and the
direction opposite that of the boost from the $B$ rest frame.  The
function $\Pi(s)$ parameterizes the $\rho$ resonance shape.  It is
defined to give the correct threshold phase-space behavior and to
incorporate the measured $\rho$ width, variations in the
parameterization of this function are one of the sources of residual  
theoretical uncertainty of this analysis.

\begin{figure}[htb]
\begin{center}
\leavevmode
\psfig{figure=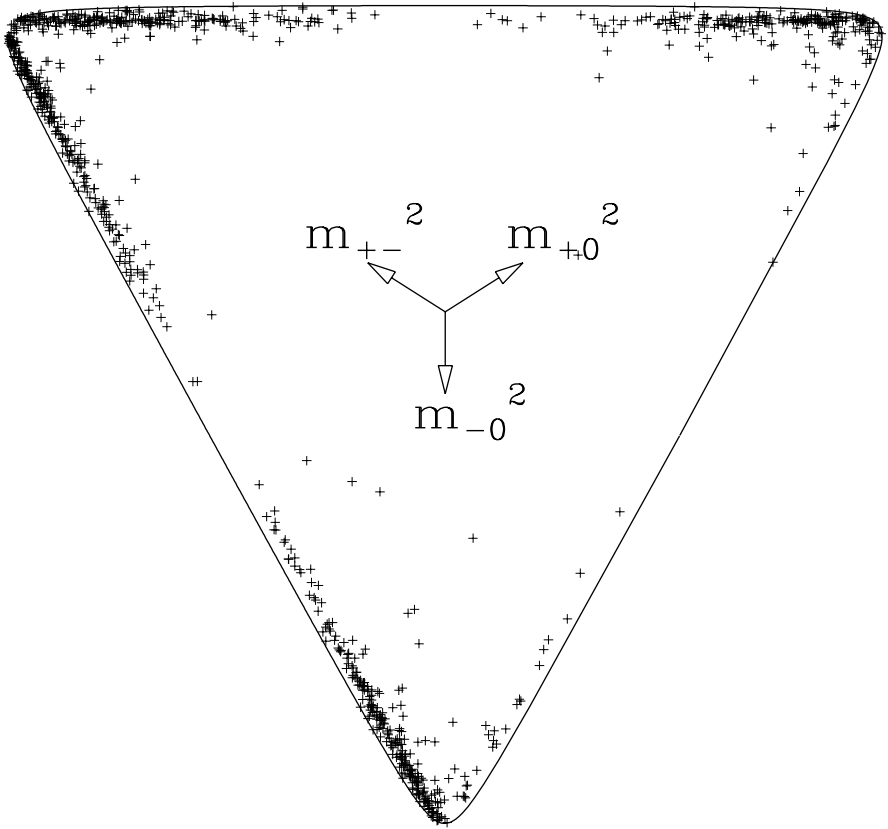,angle=90}
\end{center}
\caption[*]{The $\rho\pi$ contributions to the Dalitz plot for $B 
\rightarrow \pi^+\pi^-\pi^0$ .}
\label{fig2}
\end{figure}

The angular dependence is that associated with the decay of a
longitudinally-polarized $\rho$ meson of charge $(i+j)$ to two pions.  The
related amplitude for the $\bar B_d$ decay also contributes to the
time-dependent rate for the decay of an initially pure $B_d$ or $\bar
B_d$ state.  Interference between the different $\rho$ bands is enhanced by
the fact that the $\rho$ is longitudinally polarized and thus the
cos($\theta$) form for its decay throws the events towards the corners
of the Dalitz plot.  This is seen in Fig. \ref{fig2}, which is taken from the
BaBar Book and represents a simulation using amplitudes calculated from
a particular model.\cite{jerome}

\vspace{.5cm}

The large strong phases from the resonant behavior and the interference
of the different charge-channel contributions enhances the CP-violating
asymmetry  in the regions of the
time-dependent Dalitz plot.  A multiparameter maximum-likelihood fit to
the broad $\rho$-band regions of the time-dependent Dalitz plot is made,
with each tree and penguin amplitude parameterized by an arbitrary
magnitude and strong phase, and with the weak phases as given by the
Standard model.  The asymmetries then depend only on one combination of
weak phases, $\alpha= \pi-\beta-\gamma$ along with nine other parameters
(the magnitude and strong phases of each of the five isospin amplitudes
minus one irrelevant overall strong phase).  In principal, provided the
$\rho^0 \pi^0$ contribution is large enough, this fit will allow one to
extract not only a value of $\sin(2 \alpha$) free of uncertainties due to
penguin contributions, but also $\cos(2 \alpha$), thereby removing some of
the discrete ambiguities in the solution for the Unitarity triangle.  In
a realistic study additional parameters and assumptions must be made to
parameterize non-resonant $B$ decays to three pions and also any other
resonances that contribute significantly to the three-pion final state.
It remains to be seen whether sufficient data can be collected to make
this analysis effective when all the contributing channels and
background contributions are taken into account.  Certainly it will not
be easy.  It will require many years of data taking at a $B$ factory.
Because the final state contains a $\pi^0$ this mode is not accessible
to the current TeVatron experiments.  Preliminary studies for dedicated
hadron collider $B$ experiments suggest this mode may possibly be
feasible for study, but further work on signal to background ratios is
needed.  I still hope that this mode can eventually give us a clean
$\alpha$ measurement, but I recognize that the experimental challenge is
significant.  Some theoretical uncertainties in the value of $\alpha$
extracted in this way remain, due to the contribution of QED penguins,
and also due to the assumed constant strong phases for the isospin
amplitudes and the sensitivity to the $\rho$-shape.  However these effects 
are estimated to be small.  By the time this measurement is made I expect 
that their impact will be under much better control.

Isospin breaking effects must also be considered as a source of
theoretical uncertainties when investigating these modes.  The dominant
correction comes from the fact that, due to isospin breaking of the
quark masses, the physical $\pi^0$ and $\rho^0$ states each have a small
admixture of the isospin zero quark combination.  The consequence of
this effect is largest in the $\pi\pi$ analysis as it reintroduces the
$I_f=1$ amplitude that is otherwise forbidden by Bose statistics.  For
the $\rho\pi$ channel the impact of isospin breaking has been estimated
to be small.

\subsection{SU(3) in $K \pi$ and $\pi\pi$ and limits on $\gamma$}

Here I will briefly describe an analysis to extract the Unitarity
triangle angle $\gamma$ from the data on various channels for $B\ra K
\pi$ and $B\ra \pi\pi$.  The work I will discuss is that of Neubert and
Rosner,\cite{neubertrosnerkpi} and the subsequent paper of Neubert.
\cite{neubertkpi}   
This analysis provides an interesting example because it
uses essentially the entire toolkit of methods, the Operator Product
Expansion, diagrammatic classification of contributions, isospin and
SU(3), and finally factorization approximation as a way to estimate
SU(3) breaking corrections.  However a careful selection of the
quantities for which the least accurate approximations are used leads to
a relatively small theoretical uncertainty for the final result.  The
simple rule of thumb is that the tool with large fractional uncertainty
should, if possible, be restricted to determining a small part of the
overall result.

The decays $B\ra K\pi$ are interesting because the tree contributions
$b\ra su\bar u$ are Cabibbo suppressed.  In fact it appears that the rate
is dominated by the QCD penguin contributions.  However, as in the $\pi
\pi$ case, certain isospin channels do not have any such contribution.
The quark transition $b \ra u \bar u s$ can have $\Delta I =0, 1$ and
thus with the spectator quark added $I_f = 1/2$  or $3/2$.  The gluonic
penguin contributes only to $\Delta I =0, I_f=1/2$.  Here electroweak
penguin contributions cannot be ignored, as they enter at approximately
the same level as the Cabibbo-suppressed tree contributions, and for all
isospin amplitudes.  A major part of the work then comes in estimating
the corrections due to electroweak penguin effects, and the uncertainty
on these corrections.

The key to the analysis is to recognize that the $I_f =3/2 $ arises only
from tree diagrams and electroweak penguins.  The key initial
observation is that, in terms of the isospin-based amplitudes $A_{\Delta
I, I_f}$
\begin{eqnarray}
A(B^+ \rightarrow \pi^+ K^0) &=& A_{0,1/2} +A_{1,1/2}
+ A_{1,3/2}\nonumber\\
-\sqrt2 A(B^+ \rightarrow \pi^+ K^0) &=& A_{0,1/2}
+A_{1,1/2} - 2 A_{1,3/2}\ .
\end{eqnarray}
Gluonic penguin diagrams contribute only to $A_{0,1/2}$.
Neubert and Rosner define the following quantities
\begin{equation}
R_*  =  { Br(B^+\rightarrow K^0\pi^+) + Br(B^-\rightarrow \bar K^0\pi^-)
         \over 2( Br(B^0\rightarrow K^0\pi^+)
+ Br(\bar B^0\rightarrow  K^-\pi^0))} = (1-\Delta_*)^2 \ .
\end{equation}

One can  make the weak and strong phase dependence  explicit by writing
\begin {equation}
 A(B^+\rightarrow K^0\pi^+) = Pe^{i\phi_P}(e^{i\pi}
+ e^{i\gamma}e^{i\eta}\epsilon_a )
\end{equation}
where $\phi_P$ and $\eta$ are strong phases and, in terms of the diagrams,
\begin{eqnarray}
P& = & |\lambda_c (P_c-P_t- 1/3P_{EW,t})|\nonumber\\
\epsilon_a & = & 7|\lambda_u (P_u-P_c-A)|/P \ .
\end{eqnarray}
Similarly one can write the ratio
\begin{equation}
-3A_{1,3/2} /P = \epsilon_{3/2}e^\phi_{3/2}(e^{i\gamma} +qe^{i\omega})
\end{equation}
expanded so that the weak and strong phase structure of each term is
made explicit.  The notation is chosen so that the quantities $P$, $q$,
$\epsilon_a$ and $\epsilon_{3/2}$ are real and all phases are explicit.
Here $q e^{i\omega}$ is the ratio of electroweak penguin type
contributions to the tree type contributions to $A_{3/2}$.  Only the
top-type diagram gives a significant electroweak penguin contribution
and that enters with a coefficient $\lambda_t =-\lambda_c -\lambda_u$
but the $\lambda_u$ contribution is dropped in the above as it is too small
to matter here.

I find it convenient to introduce the quantities
\begin{eqnarray}
r_{3/2} &=& {-3A_{1,3/2} \over A(B^+\rightarrow K^0\pi^+)} 
= {- \epsilon_{3/2}e^{\phi_{3/2}-\phi_P}(e^{i\gamma} +qe^{i\omega})
\over 1- e^{i\gamma}e^{i\eta}\epsilon_a }\nonumber\\[1.5ex]
\xi &=& {1-a(0+)\over 1+a(0+)}
= \left|{A(B^-\rightarrow \bar K^0\pi^-)\over A(B^+\rightarrow K^0\pi^+)}\right|^2 \ .
\end{eqnarray}
Then one can write
\begin{equation}
R_*^{-1} = {|1 + r_{3/2}|^2 +\xi|1 +\bar r_{3/2}|^2 \over 1 +\xi}\ .
\end{equation}
In the above equations the CP-conjugated amplitudes are obtained from
their CP partners by simply changing the sign of the weak phase $\gamma$
everywhere (since $e^{i\pi} = e^{-i\pi}$).

A major point of introducing all this notation is that the quantities
$\epsilon_a$, $\epsilon_{3/2}$ and $q e^{i\omega}$ are all small, the
first two because they are suppressed by the ratio
$|\lambda_u/\lambda_c|$ and the last because it is a ratio of
electroweak penguin to tree, albeit enhanced by the inverse CKM ratio
$|\lambda_c/\lambda_u|$.  Useful results can be obtained keeping only
the leading effects of these quantities.  Relatively large uncertainties
in these quantities translate into only small uncertainties in $R_*$.

This statement (which is mine, not Neubert's) is a bit of a cheat, since
the sensitivity to $\gamma$ is not in the value of $R_*$ but in its
deviation from 1, which is expected to be small for the same reason.
The interest in this problem is sparked by preliminary data from CLEO
which give $R_* = 0.47\pm 0.27$.  If the value of $R_*$ deviates
significantly from 1 then the above equations can be used to put
interesting constraints on the allowed range of gamma, provided we can
constrain the quantities $\epsilon_a$, $\epsilon_{3/2}$ and $q
e^{i\omega}$.  The better we can constrain these parameters, the more
likely we are to be able to determine whether beyond Standard Model
physics is needed to explain the measurement.  Further we will need some
information on strong phase differences.  However even generous ranges
on these quantities may translate into constraints on the allowed range
of gamma.  So now let us pursue the question of how and how well we can
calculate each of these quantities.

The quantities $qe^{i\omega}$ turns out to be cleaner than one would
expect.  In general two operators contribute for the tree amplitude and
four for the electroweak penguin.  However two of these  latter four give very
small contributions to this matrix element and can be neglected.  The
other two are Fierz-equivalent to the two tree-type operators.
Furthermore only one linear combination of these two operators
contributes in the SU(3) limit, the matrix element of the other must
vanish.  This is another application of Bose statistics, this time to
the U-spin part of SU(3).  Thus even though $qe^{i\omega}$ is a ratio of
an electroweak penguin amplitude to a tree-type amplitude each is
dominated by a single operator in the SU(3) limit. Furthermore and the
two operators (for the two diagrams) are Fierz-equivalent to
one-another.  This means that only a single strong phase enters---the
same for both these contributions, so that the ratio is fixed by the
ratio of coefficients in this limit.  Thus, Neubert writes $qe^{i\omega} = (1-\kappa e^{i\Delta_{3/2}})\delta_+$ where $\delta_+$ is given
by the ratio of coefficients of the $I=3/2$ electroweak and tree
operators that survive in the SU(3) limit and $\kappa e^{i\Delta_{3/2}}$ is
the SU(3) breaking correction to this quantity.  One can then estimate
such corrections and the uncertainties in them.  First one
estimates SU(3) breaking correction $\kappa$ by calculating it in the
factorization approximation.  Neubert estimates this effect to be
$(6\pm6)\%$.  In this approximation $\Delta_{3/2}=0$.  This then gives
$q e^{i\omega} \approx \delta_{EW} =(1-\kappa)\delta_+= 0.64 \pm 0.15$,
where the large percentage error reflects the large theoretical
uncertainties inherent in the SU(3) and factorization approximation as
well as the smaller but still significant uncertainty in the evaluation
of the ratio of operator coefficients that reflects small residual scale
and scheme dependence of this ratio.  He also includes the effect of
allowing non-zero $\Delta_{3/2}$ values in this overall error
estimation, noting that allowing a phase $|\Delta_{3/2}|\le 90^\circ$
would yield only $|\omega|\le 2.7^\circ$.

For the quantity $\epsilon_{3/2}$ one must again rely on SU(3), which
relates the $B^+ \ra K \pi, I=3/2$ tree amplitude to the corresponding
tree amplitude for $B^+ \ra \pi \pi, I=2$.  The measured charged $B \ra
\pi K$ rates determines the magnitude of penguin amplitude in the
denominator of epsilon, up to corrections of order $\epsilon_a $ which
we will discuss below.  Here one expects a large SU(3) correction.  This
is estimated again by calculating the correction in the factorization
limit, taking the factorization model parameters $a_1^{ij}$ and
$a_2^{ij}$ (where $ij = K\pi$ or $\pi\pi$) and the ratio $f_K/f_\pi$ from
fits to data.  The only model dependent part of this SU(3) correction
calculation is the ratio $F(B\ra K)/F(B\ra \pi)$ which is 1 in the SU(3)
limit.  Models all agree with the range $1.1\pm 0.1$.  Since this factor
enters the $\epsilon_{3/2}$ factorization calculation with a relatively
small coefficient, the impact of its large uncertainty on the overall correction 
factor is not great.  Again one must assign some
uncertainty to the difference between the factorization-model based
estimate of the SU(3) correction and the actual SU(3) breaking effects,
but it is reasonable to expect that this estimate has correctly
accounted for the largest part of SU(3) breaking corrections.  Including
this and all the various sources of uncertainty, both theoretical and
experimental, Neubert estimates about a $25 \%$ uncertainty in the
extracted value of $\epsilon_{3/2}$.

The remaining quantity $\epsilon_a$ is inherently small because it a
ratio of Cabibbo-suppressed to Cabibbo-allowed terms.  It would be a
source of direct CP violation $\xi \neq 1$ and may eventually be
constrained by measurement of the CP-asymmetry in $B^\pm\rightarrow K^\pm
\pi^0$ decays.  Another constraint comes from using SU(3) to relate
these decays to the $B^\pm\rightarrow K^\pm K^0$ (or $\bar K^0)$ decays.  (For an
alternate discussion of uncertainty introduced by this approach see
M. Gronau and D. Pirjol 
\cite{gronaukpi}).  Further, $\epsilon_a$ can be re-expressed in terms
of a difference of $I=1/2$ and $I=3/2$ amplitudes that arises solely due
to rescattering effects.  Neubert uses all of these arguments to
estimate a ``reasonable'' and a ``conservative'' (which in this context
means a more generous) range for this quantity and then explores how the
constraints on gamma vary as one varies $\epsilon_a$ over these ranges.

My point in describing this calculation is not to present the results,
which you can read in Neubert's paper, and which indeed will change with
time as experimental numbers improve.  What I want to show is how the
tools of SU(3) limit and factorization can be combined to obtain results
which are better than either tool used separately.  First the SU(3)
limit prediction is calculated.  Then the correction to that limit is
calculated using the factorization approximation.  Thus the uncertainty
from factorization in the result is the uncertainty in the correction to
SU(3) rather than the uncertainty in the entire effect.  This is clearly
an improvement over a straightforward use of either uncorrected SU(3) or
simple factorization estimates to calculate the entire effect.

Even when such tricks are used to the full extent available still the
question remains: how big is the uncertainty in the result after all
is said and done?  Unfortunately the answer is never clean.  But clearly
the problem is much reduced if we are debating whether an effect is
$6\%$ or twice as big rather than whether it is $50\%$ or twice that.
The challenge to theorists is to make the sources of their uncertainties
clear, and to do as honest a job as possible of constraining them.
Here work remains to be done.  Neubert's paper gives an example of a
serious attempt to explore such questions in a systematic way, for a
particular set of decays.  

\subsection{Theoretical Uncertainties}

In the end, whatever the estimates might be,
it is important to remember that theoretical uncertainty is not
statistical; it is simply wrong to talk about the probabilities of
certain results as if these estimates were in fact gaussian-based
standard deviations.  It is also very misleading to combine different
sources of theoretical error by adding them in quadrature, though one
sees this done frequently in the literature.

A false division between theoretical uncertainties and systematic errors in an
experimental value is often made---at least in the minds of theorists
making the initial predictions.  A theorist makes a clean prediction
with small theoretical errors for a quantity---say, for example, the CP-violating 
asymmetry in inclusive $b\ra u\bar u d$ decays.  The theorist
is happy.  However that quantity is in fact impossible to measure, since
any real experiment has aperture limitations and in addition must apply
cuts to separate the signal from background, in the example above both
that from sources other than $B$-decays and that from the dominant $b\ra
c \bar q q'$ decays.  The impact of these cuts on the relationship of
the measurement to the prediction must be evaluated based on some
theoretical models.  This is where the large theoretical errors will
typically appear.  

Experimentalists now often quote their uncertainties by
separating out such effects as theoretical uncertainties rather than by
including them in the overall systematic uncertainties.  
My point here is that the magnitude of this theoretical uncertainty typically will
have nothing to do with the magnitude of the theoretical uncertainty for
this measurement given in the original theoretical predictions.  Such
experiment-dependent theoretical uncertainties belong neither to the
domain of pure theory nor to the domain of experiment, but live at the
interface between them.  They do, however, suffer the usual disease of
theoretical errors---they are not statistical effects.  It would be very
helpful if theorists making their clean predictions could at least
consider and briefly discuss what impact experimental cuts will have on
the validity of their prediction.  I do not mean the theorist should define 
specific cuts, but rather should discuss the
question of whether the result can survive any cuts at all without serious
degradation.

My remarks above are borne out in a well-known way in the case of the
extraction of the magnitude of the CKM-parameter $V_{ub}$ from
semileptonic $B$-decay data.  Two general classes of methods are in the
market---one uses exclusive decays and has obvious theoretical
uncertainties related to the form factors (that is the QCD matrix
elements) that govern the particular decay in question.  The other uses
inclusive semi-leptonic decays and is thus at first sight very clean.
But the experiment must make cuts to remove the $b\ra cl\nu$ backgrounds.
The prediction for the cut data sample has comparable theoretical
uncertainties to the exclusive decay cases.  Eventually we need to
explore both kinds of methods, since the theoretical uncertainties for
the two approaches are essentially different.

Recently new predictions for extracting this same parameter from
hadronic measurements have appeared.  Again one group of theorists
advocates an inclusive approach, and others advocate certain exclusive
channels.  Both are interesting; both will probably have significant
theoretical errors once the real experimental limitations on the
inclusive methods are understood.

In all these cases, whether semi-leptonic or hadronic decays are considered,
one cannot use any of the more rigorous tools discussed above to
estimate the theoretical uncertainties introduced due to experimental
cuts or those due to form-factor estimates.  One is forced to resort
to models.  Often the models work at the quark rather than the hadron
level and then apply the notion of quark-hadron duality which is the
assumption that the two-body hadron kinematics reflects the underlying
quark kinematics.  This is called ``local quark-hadron duality''.  It is
not a justifiable assumption.

Estimates of theoretical errors in such cases tend to be very
subjective.  There really is no clean way to obtain them.  The most
common method is to try a few models and take the range of the results
as the range of theoretical uncertainties.  This is risky, since all the
models on the market may contain the same unjustified assumption (for
example that a particular form factor can be parameterized as a simple
pole).  Nonetheless it is common practice and perhaps the best we can
do.  My advice is one should simply be aware when this is the nature of
the theoretical error estimate and treat the resulting numbers with a
sufficient amount of salt.  The recent history of statements about
errors in estimates of $\epsilon^\prime/\epsilon$ should be a clear
object lesson to experimenters on the reliability of theoretical error
estimates.

\section*{Acknowledgments}

I would like to thank Jo\~ao Silva  for a careful reading of these lectures
and his helpful suggestions for improvements.

 \end{document}